\documentclass[12pt]{article}

\pdfoutput=1

\usepackage{graphics}
\usepackage{epsf}
\usepackage{epsfig}
\usepackage{graphicx}
\usepackage{amsfonts}
\usepackage{amssymb}
\usepackage{amsmath}
\usepackage{amsthm}
\usepackage{pifont}
\usepackage{multirow}
\usepackage{latexsym}
\usepackage{verbatim}
\usepackage{mcite}
\usepackage{cite}
\usepackage{units}
\usepackage[all]{xy}
\usepackage{cancel}
\usepackage{slashed}
\usepackage{a4wide}
\usepackage{hyperref}

\usepackage{fancyvrb}
\DefineVerbatimEnvironment{code}{Verbatim}{fontsize=\scriptsize}

\def\a{{\alpha}}
\def\b{{\beta}}

\def\g{{\gamma}}

\def\m{{\mu}}

\def\bfone{\relax{\rm 1\kern-.35em 1}}

\newcommand{\cN}{{\cal N}}

\newcommand{\be}{\begin{equation}}
\newcommand{\ee}{\end{equation}}
\newcommand{\ben}{\begin{displaymath}}
\newcommand{\een}{\end{displaymath}}
\newcommand{\bea}{\begin{eqnarray}}
\newcommand{\eea}{\end{eqnarray}}

\newcommand{\bean}{\begin{eqnarray*}}
\newcommand{\eean}{\end{eqnarray*}}

\DeclareMathAlphabet{\mathpzc}{OT1}{pzc}{m}{it}

\begin{document}
\pagestyle{plain}


\makeatletter \@addtoreset{equation}{section} \makeatother
\renewcommand{\thesection}{\arabic{section}}
\renewcommand{\theequation}{\arabic{equation}}
\renewcommand{\thefootnote}{\arabic{footnote}}


\setcounter{page}{1} \setcounter{footnote}{0}


\begin{titlepage}

\begin{flushright}
UUITP-03/13\\
\end{flushright}

\bigskip

\begin{center}

\vskip 0cm

{\LARGE \bf Fully stable dS vacua from generalised fluxes}\\[6mm]

\vskip 0.5cm

{\bf Johan Bl{\aa}b\"ack, Ulf Danielsson \,and\, Giuseppe Dibitetto}\\

\vskip 25pt

{\em Institutionen f\"or fysik och astronomi, \\ 
Uppsala Universitet, \\ 
Box 803, SE-751 08 Uppsala, Sweden \\
{\small {\tt \{johan.blaback, ulf.danielsson, giuseppe.dibitetto\}@physics.uu.se}}} \\

\vskip 0.8cm

\end{center}

\vskip 1cm

\begin{center}

{\bf ABSTRACT}\\[3ex]

\begin{minipage}{13cm}
\small

We investigate the possible existence of (meta-)stable de Sitter vacua within $\cN=1$ compactifications with generalised fluxes. With the aid of an algorithm inspired by the method of differential
evolution, we were able to find three novel examples of completely tachyon-free de Sitter extrema in a non-isotropic type IIB model with non-geometric fluxes. 
We also analyse the surroundings of the aforementioned points in parameter space and chart the corresponding stability regions. These happen to occur at small values of the cosmological constant compared to 
the AdS scale. 

\end{minipage}

\end{center}

\vfill

\end{titlepage}



\section*{Introduction}

The possible existence of (meta-)stable de Sitter (dS) vacua in string
theory, is considered to be of crucial 
importance if we want to achieve agreement with the main cosmological
observations 
of the last two decades. The problem has been addressed following various
approaches using string compactifications, and their link with the
corresponding effective descriptions. Focusing on particular effective
supergravity descriptions, which often are referred to as $STU$-models, the
introduction of fluxes turns out to be the key, at least perturbatively, for
inducing a superpotential of the right form 
\cite{Giddings:2001yu}. 

Within these supergravity models, several theories have been studied that
describe string backgrounds in type IIA 
\cite{Kachru:2002he, Derendinger:2004jn, DeWolfe:2004ns, Villadoro:2005cu,
Derendinger:2005ph, DeWolfe:2005uu, 
Flauger:2008ad, Caviezel:2008tf, Danielsson:2009ff, Danielsson:2010bc,
Danielsson:2011au, Danielsson:2012et}, 
and type IIB \cite{Camara:2005dc, Aldazabal:2006up, Aldazabal:2007sn,
Guarino:2008ik, deCarlos:2009fq, deCarlos:2009qm}. So far, there are no
known example of a fully meta stable dS vacuum. All geometric
compactifications have unstable directions 
already within the isotropic truncation of the $STU$-model. Using
non-geometric fluxes \cite{Shelton:2005cf} it was possible to construct the
first example of an isotropically stable dS extremum \cite{deCarlos:2009fq}.
This solution is still, however, unstable in the non-isotropic directions.
This situation was recently analysed in a more systematic way in 
ref.~\cite{Danielsson:2012by}. 

The main result of this paper are the first examples of fully stable dS
vacua in the $STU$-model. We show that non-geometric fluxes provide enough
freedom in parameter space for tuning the masses of all the moduli to be 
positive, even the non-isotropic ones. In appendix~\ref{app:fluxfig}, we
analyse the whole region of stable dS in parameter space 
around such stable dS points by providing some insightful plots. In
appendix~\ref{app:algorithm}, we give some details concerning the algorithm
that allowed us to find the aforementioned stable dS vacua.

\section*{The $\mathbb{Z}_{2}\,\times\,\mathbb{Z}_{2}$ orbifold with generalised fluxes}

Orbifold compactifications of type IIB string theory on $T^{6}/(\mathbb{Z}_{2}\,\times\,\mathbb{Z}_{2})$ with O3/O7-planes (and duals thereof) with generalised fluxes can all be placed within the same framework
of effective four-dimensional supergravity descriptions that are known as $STU$-models. These theories enjoy $\cN=1$ supersymmetry and $\textrm{SL}(2)^{7}$ global bosonic symmetry.
The action of such a global symmetry on the fields and couplings can be interpreted as the effect of string dualities. 

The scalar sector contains seven complex fields spanning the coset space $\left(\textrm{SL}(2)/\textrm{SO}(2)\right)^{7}$ which we denote by $\Phi^{\a}\,\equiv\,\left(S,T_{i},U_{i}\right)$ with $i=1,2,3$. 
The kinetic Lagrangian follows from the K\"ahler potential
\be
\label{Kaehler_STU}
K\,=\,-\log\left(-i\,(S-\overline{S})\right)\,-\,\sum_{i=1}^{3}{\log\left(-i\,(T_{i}-\overline{T}_{i})\right)}\,-\,\sum_{i=1}^{3}{\log\left(-i\,(U_{i}-\overline{U}_{i})\right)}\ .
\ee
This yields
\be
\mathcal{L}_{\textrm{kin}} = \frac{\partial
S\partial \overline{S}}{\left(-i(S-\overline{S})\right)^2} + \, \sum_{i=1}^{3}\left(\frac{\partial
T_{i}\partial \overline{T}_{i}}{\left(-i(T_{i}-\overline{T}_{i})\right)^2} + \frac{\partial
U_{i}\partial \overline{U}_{i}}{\left(-i(U_{i}-\overline{U}_{i})\right)^2}\right) \ .
\ee

The presence of fluxes induces a scalar potential $V$ for the moduli fields which is given in terms of the above K\"ahler potential and a holomorphic superpotential $W$ by
\be
\label{V_N=1}
V\,=\,e^{K}\left(-3\,|W|^{2}\,+\,K^{\a\bar{\b}}\,D_{\a}W\,D_{\bar{\b}}\overline{W}\right)\ ,
\ee
where $K^{\a\bar{\b}}$ is the inverse K\"ahler metric and $D_{\alpha}$ denotes the K\"ahler-covariant derivative.\footnote{We are working in units where $M_{\textrm{Pl}}=1$.}

The general form of a superpotential induced by locally geometric fluxes in type IIB with O3/O7-planes is given by
\be
\label{W_Complete}
W\,=\underbrace{\,P_{F}(U_{i})\,}_{F \textrm{ flux}}\,+\,S\,\underbrace{\,P_{H}(U_{i})\,}_{H \textrm{ flux}}\,+\,\sum\limits_{k}{T_{k}\,\underbrace{\,P_{Q}^{(k)}(U_{i})\,}_{Q \textrm{ flux}}}\,+\,S\,\sum\limits_{k}{T_{k}\,\underbrace{\,P_{P}^{(k)}(U_{i})\,}_{P \textrm{ flux}}}\ ,
\ee
where $P_{F}$, $P_{H}$, $P_{Q}^{(k)}$ and $P_{P}^{(k)}$ are cubic polynomials in the complex structure moduli given by
\be
\begin{array}{cclc}
P_{F}(U_{i}) & = & a_{0}\,-\,\sum\limits_{i}{a_{1}^{(i)}\,U_{i}}\,+\,\sum\limits_{i}{a_{2}^{(i)}\,\dfrac{U_{1}\,U_{2}\,U_{3}}{U_{i}}}\,-\,a_{3}\,U_{1}\,U_{2}\,U_{3} & , \\[3mm]
P_{H}(U_{i}) & = & -b_{0}\,+\,\sum\limits_{i}{b_{1}^{(i)}\,U_{i}}\,-\,\sum\limits_{i}{b_{2}^{(i)}\,\dfrac{U_{1}\,U_{2}\,U_{3}}{U_{i}}}\,+\,b_{3}\,U_{1}\,U_{2}\,U_{3} & , \\[3mm]
P_{Q}^{(k)}(U_{i}) & = & c_{0}^{(k)}\,+\,\sum\limits_{i}{c_{1}^{(ik)}\,U_{i}}\,-\,\sum\limits_{i}{c_{2}^{(ik)}\,\dfrac{U_{1}\,U_{2}\,U_{3}}{U_{i}}}\,-\,c_{3}^{(k)}\,U_{1}\,U_{2}\,U_{3} & , \\[3mm]
P_{P}^{(k)}(U_{i}) & = & -d_{0}^{(k)}\,-\,\sum\limits_{i}{d_{1}^{(ik)}\,U_{i}}\,+\,\sum\limits_{i}{d_{2}^{(ik)}\,\dfrac{U_{1}\,U_{2}\,U_{3}}{U_{i}}}\,+\,d_{3}^{(k)}\,U_{1}\,U_{2}\,U_{3} & .
\end{array}
\ee
The IIA and IIB interpretation of the above superpotential couplings is summarised in table~\ref{table:unprimed_fluxes}. Note that by having the possibility of choosing T-duality frame, observables that are not invariant under T-duality will differ depending on what frame is chosen.
 
\begin{table}[h!]
\renewcommand{\arraystretch}{1.25}
\begin{center}
\scalebox{0.92}[0.92]{
\begin{tabular}{ | c || c | c | c | c |}
\hline
couplings & Type IIB & Type IIA & fluxes & \emph{dof}'s\\
\hline
\hline
$1 $&  $ {F}_{ mnp} $& $F_{ambncp}$ & $  a_0 $ & $1$\\
\hline
$U_{i} $&  ${F}_{ mn c} $& $F_{ambn}$ & $   -a_1^{(i)} $ & $3$\\
\hline
$U_{j}U_{k} $& ${F}_{m b c} $& $F_{am}$ & $  a_2^{(i)} $ & $3$\\
\hline
$U_{i}U_{j}U_{k} $& ${F}_{a b c} $& $F_{0}$ & $  -a_3 $ & $1$\\
\hline
\hline
$S $& $ {H}_{mnp} $& $ {H}_{mnp} $  & $  -b_0$ & $1$\\
\hline
$S \, U_{i} $& ${H}_{mn c} $& ${{\omega}_{mn}}^{c}$ & $  b_1^{(i)} $ & $3$\\
\hline
$S \, U_{j}U_{k} $&  ${H}_{ m b c}$ & $ {{Q}_{ m }}^{ b c}$  & $  -b_2^{(i)} $ & $3$\\
\hline
$S \, U_{i}U_{j}U_{k} $& $ {H}_{a b c} $& $ {R}^{a b c} $ & $  b_3 $ & $1$\\
\hline
\hline
$T_{i} $& $  {Q_p}^{a b} $&$ H_{a b p} $ & $  c_0^{(i)} $ & $3$\\
\hline
$T_{i} \, U_{j} $ & $ {Q_p} ^{a n} = {Q_p}^{m b} \,\,\,,\,\,\, {Q_a}^{b c} $& $ {\omega_{p a}}^{n} = {\omega_{b p}}^{m} \,\,\,,\,\,\, {\omega_{b c}}^a $  & $c_1^{(ji)} $ & $9$\\
\hline
$T_{l} \, U_{i}U_{j}$ & $ {Q_c}^{mb} = {Q_c}^{a n} \,\,\,,\,\,\, {Q_p}^{mn} $& $ {Q_b}^{cm} = {Q_a}^{n c} \,\,\,,\,\,\, {Q_p}^{mn} $ & $-c_2^{(kl)} $ & $9$\\
\hline
$T_{l} \, U_{i}U_{j}U_{k} $ & $  {Q_{c}}^{mn} $& $  R^{mnc} $ & $-c_3^{(l)} $ & $3$\\
\hline
\hline
$S \, T_{i} $ & $ {P_p}^{a b}$ & & $  -d_0^{(i)} $ & $3$\\
\hline
$S \, T_{i} \, U_{j} $ & $ {P_p}^{a n} = {P_p}^{m b} \,\,\,,\,\,\, {P_a}^{b c} $&  & $-d_1^{(ji)} $ & $9$\\
\hline
$S \, T_{l} \, U_{i}U_{j} $ & $ {P_c}^{mb}= {P_c}^{a n} \,\,\,,\,\,\, {P_p}^{mn} $&  & $d_2^{(kl)} $ & $9$\\
\hline
$S \, T_{l} \, U_{i}U_{j}U_{k} $ & $  {P_{c}}^{mn} $&  & $d_3^{(l)} $ & $3$\\
\hline
\end{tabular}
}
\end{center}
\caption{{\it Mapping between fluxes and couplings in the superpotential both in type IIB with O$3$ and O$7$ and in type IIA with O$6$. The six internal directions of $T^{6}$ are split into $\,``-"$ labelled by $m=1,3,5$ and $\,``\,|\,"$ labelled by $a=2,4,6$. Note that the empty boxes in type IIA are related to the presence of dual fluxes 
which do not even admit any local description. Note that the orbifold involution forces $i,j,k$ to be all different any time they appear as indices of fields of the same type ($T$ or $U$).}}
\label{table:unprimed_fluxes}
\end{table}

\section*{Fully stable de Sitter vacua}

The setup we focus on is the IIB duality frame with O$3$- and O$7$-planes with generalised fluxes defining a locally geometric background, \emph{i.e.} $F_{3}$, $H$, $Q$ and $P$ fluxes. 
In ref.~\cite{Danielsson:2012by} it was recently argued that, whenever one includes a number of fluxes equal to twice the number $N$ of real fields in the theory, there is just enough freedom for casting the equations of 
motion into the form of a set of linear conditions, when the scalars at taken at the origin $\Phi_{0}$ of moduli space. The space of solutions of the full problem will then have dimension $N$. 
The general solution will give the fluxes as a function of the $N$ supersymmetry- (SUSY-)breaking parameters.

The SUSY-breaking parameters are $N$ real constants $A_{\a}$ and $B_{\a}$ defined through
\be
\label{SUSY_break}
D_{\a}W|_{\Phi_{0}}=A_{\a} \,+\, iB_{\a} \ .
\ee
In our non-isotropic locally geometric setup with $N=14$ real fields, one would then need to consider $28$ fluxes in order to apply the prescription of ref.~\cite{Danielsson:2012by} to the search for stable
dS vacua. The set of $28$ superpotential couplings that we have chosen here corresponds to the most general $F_{3}$ and $H$ fluxes plus the first half of $Q$ flux components in type IIB. From table~\ref{table:unprimed_fluxes},
this set reads
\be
\label{HFQ_fluxes}
\left\{a_0,\,a_1^{(i)},\,a_2^{(i)},\,a_3,\,b_0,\,b_1^{(i)},\,b_2^{(i)},\,b_3,\,c_0^{(i)},\,c_1^{(ij)}\right\}\ .
\ee

The goal of this paper is to perform a scan of the $14$-dimensional parameter space of solutions searching for fully stable dS extrema. In ref.~\cite{Danielsson:2012by} it was already argued that
a random scan would not be efficient enough for finding special tachyon-free dS extrema, since the fraction of such points in the dS region is expected to be extremely tiny.

To overcome this difficulty, we have designed an evolutionary algorithm, which has the property of flowing in parameter space towards better-behaved solutions, \emph{i.e.} positive cosmological constant and positive
eigenvalues of the mass matrix. This algorithm is presented in detail in appendix~\ref{app:algorithm}.

Letting the algorithm run with the input parameters given in appendix \ref{app:algorithm} gave rise to $3$ fully stable dS vacua, which are shown in table~\ref{dS_sols1}.
The full set of flux numbers for these solutions are specified in table~\ref{dS_sols2} in appendix \ref{app:fluxfig} .
\begin{table}[h!]
\renewcommand{\arraystretch}{1.25}
\begin{center}
\scalebox{0.90}[0.90]{
\begin{tabular}{ | l || c | c | c | }
\hline
  &  Sol.~$1$ & Sol.~$2$ & Sol.~$3$ \\
\hline 
\hline
$V_{0}\,\equiv\,V(\Phi_{0})$ & $8.85\,\times\,10^{-6}$ & $1.58\,\times\,10^{-5}$ & $1.00\,\times\,10^{-4}$\\
\hline
$\tilde{\g}\,\equiv\,\frac{|DW|^{2}}{3|W|^{2}}$ & $1.00008$ & $1.00012$ & $1.00065$\\
\hline
\begin{tabular}{c} Normalised \\[-3mm] masses \\ $(m^{2}/V_{0})$ \end{tabular}
& $\begin{array}{cc} 1.85148\,\times\,10^6 & 1.78128\,\times\,10^6 \\ 1.31064\,\times\,10^6 & 1.27212\,\times\,10^6 \\ 113890 & 94187.9 \\ 23907.3 & 11397.5 \\ 5290.32 & 1478.37 \\ 1353.35 & 607.799 \\ 17.9045 & 2.85612\,\times\,10^{-3} \end{array}$ & 
$\begin{array}{cc} 262778 & 259704 \\ 160140 & 153601 \\ 24219.8 & 11296.3 \\ 9273.53 & 7282.96 \\ 4155.21 & 1745.39 \\ 1306.57 & 343.391 \\ 24.7202 & 11.4504 \end{array}$ & 
$\begin{array}{cc} 34702.4 & 28731.5 \\ 25215.5 & 18728.6 \\ 5370.90 & 3609.62 \\ 1572.22 & 1179.86 \\ 723.060 & 518.923 \\ 188.606 & 145.959 \\ 5.16471 & 9.52228\,\times\,10^{-4} \end{array}$\\
\hline
\end{tabular}
}
\end{center}
\caption{{\it The physical quantities characterising the $3$ stable dS extrema found through the algorithm presented in appendix~\ref{app:algorithm}. The first row shows the values of the cosmological
constant, the second row the normalised energy $\tilde{\g}$, and finally the third row shows the full mass spectra normalised to the cosmological constant.}}
\label{dS_sols1}
\end{table}
%


Finally we would like to carry out a complementary analysis w.r.t. the search that we have just presented. The aim is to map out the region of stable dS inside parameter space in the 
IIB case previously introduced and studied. By means of such an analysis one can study how stable dS vacua organise themselves in the parameter space spanned by $\left\{A_{\a},B_{\a}\right\}$. 
 
\begin{figure}[h!]
\begin{center}
\scalebox{0.5}[0.5]{
\begin{tabular}{ccc}
\includegraphics[scale=1.1,keepaspectratio=true]{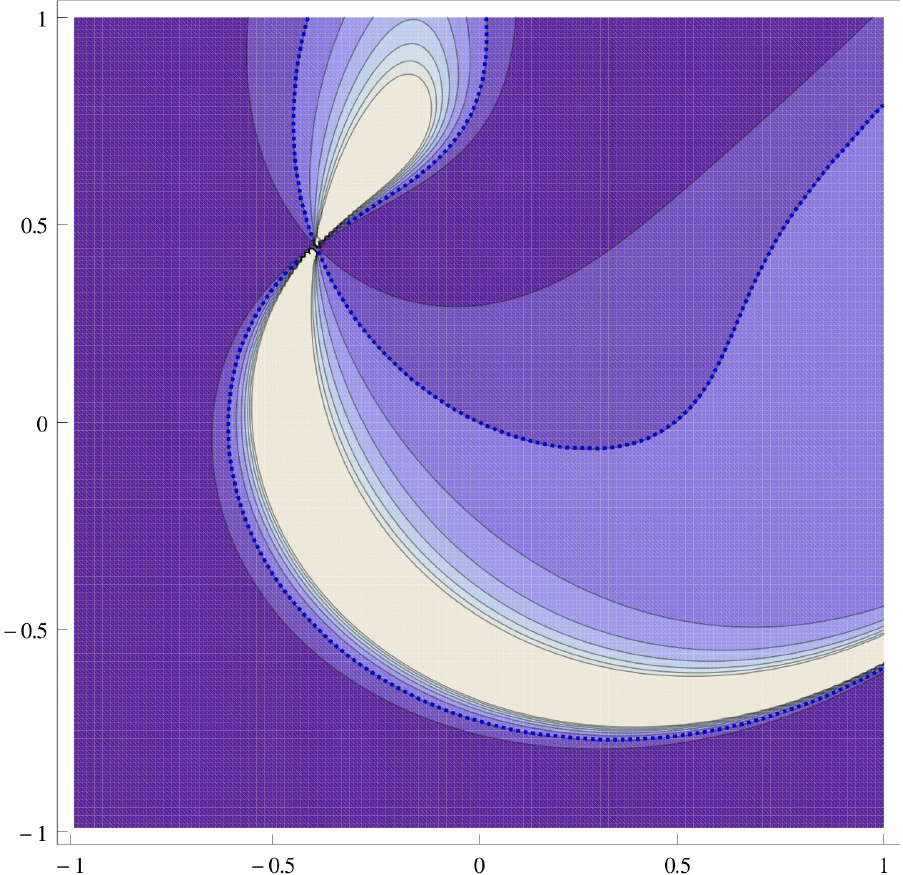} & \includegraphics[scale=1.1,keepaspectratio=true]{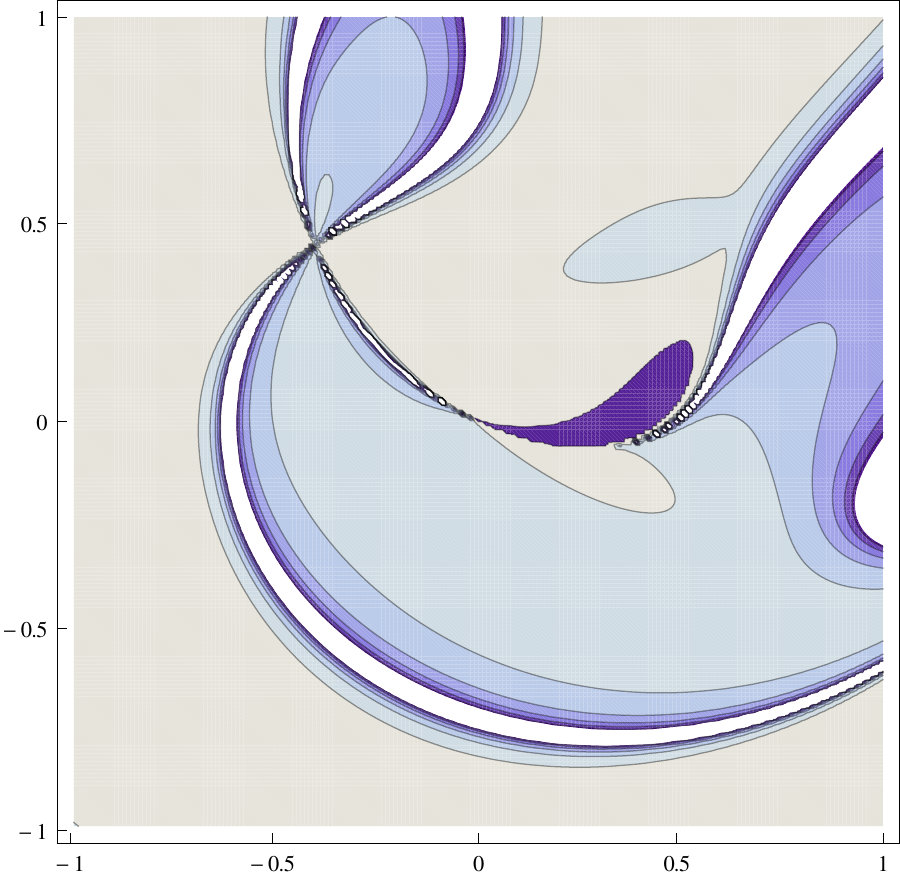} & \includegraphics[scale=1.1,keepaspectratio=true]{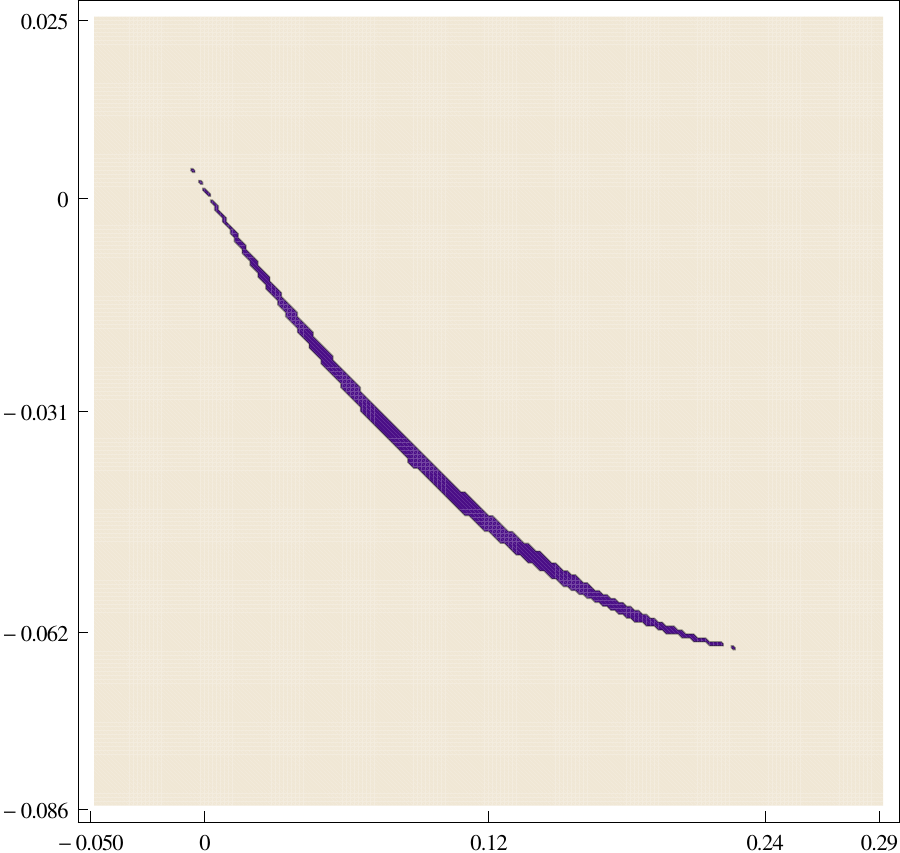}\\
\includegraphics[scale=1.1,keepaspectratio=true]{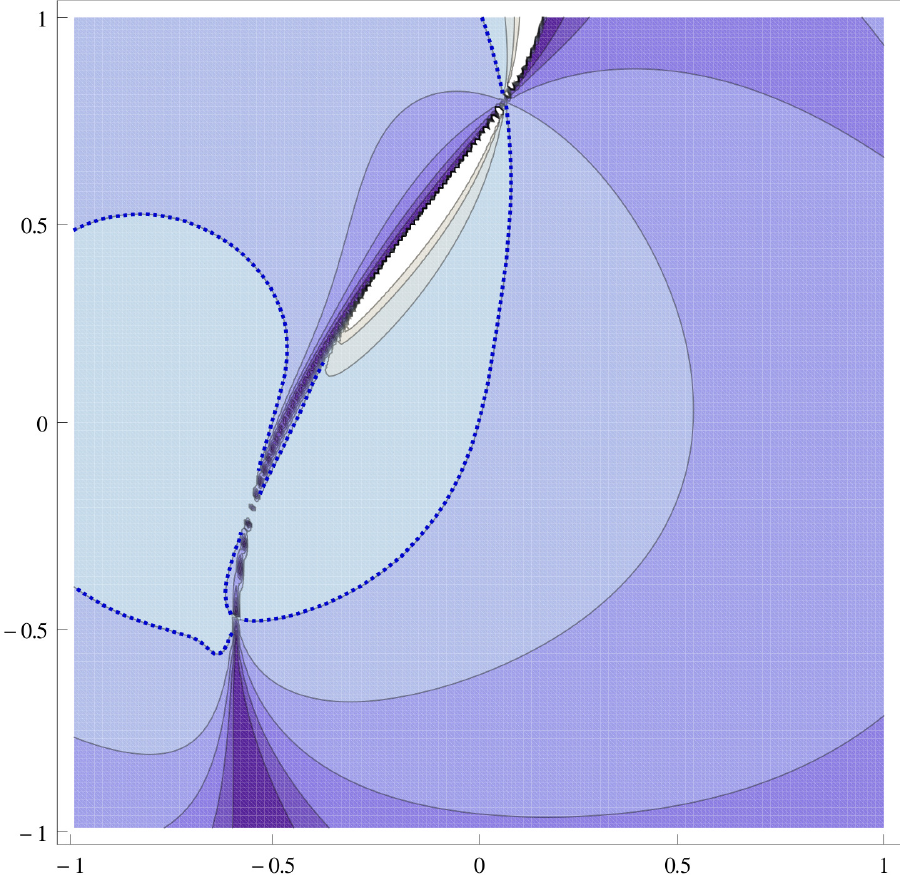} & \includegraphics[scale=1.1,keepaspectratio=true]{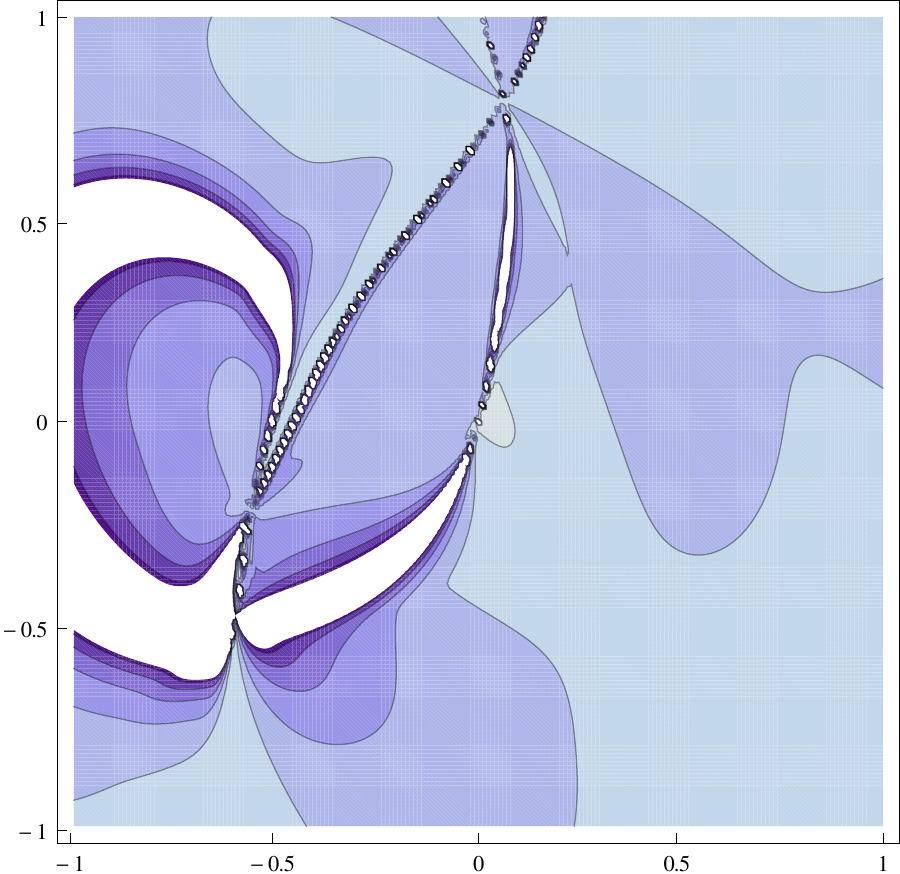} & \includegraphics[scale=1.1,keepaspectratio=true]{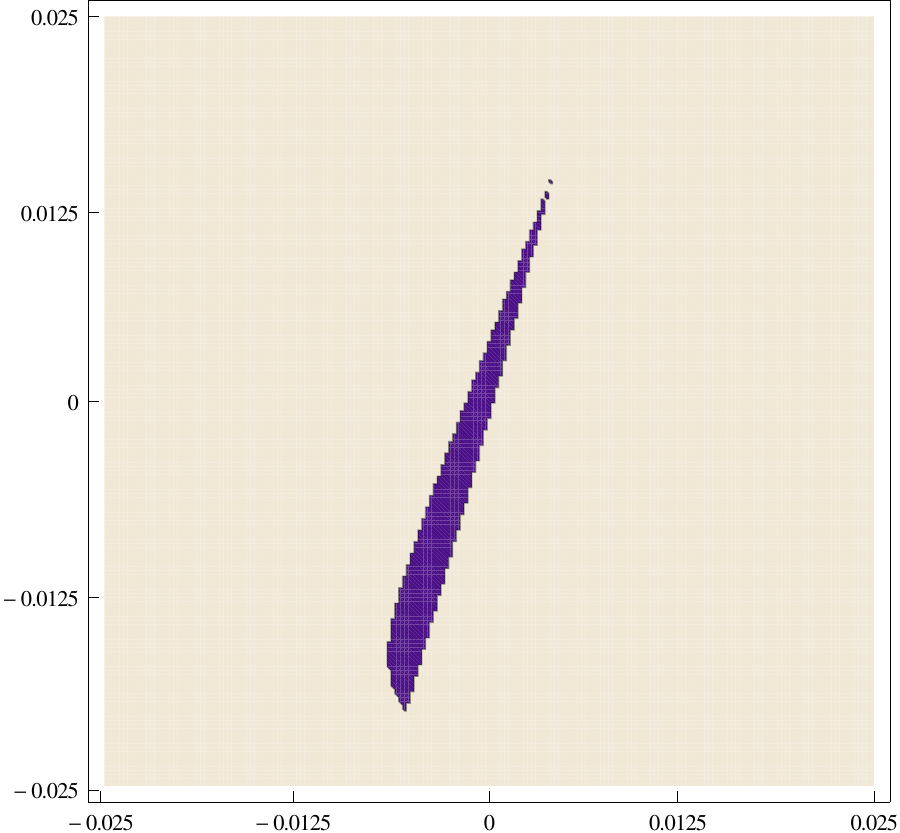}\\
\includegraphics[scale=1.1,keepaspectratio=true]{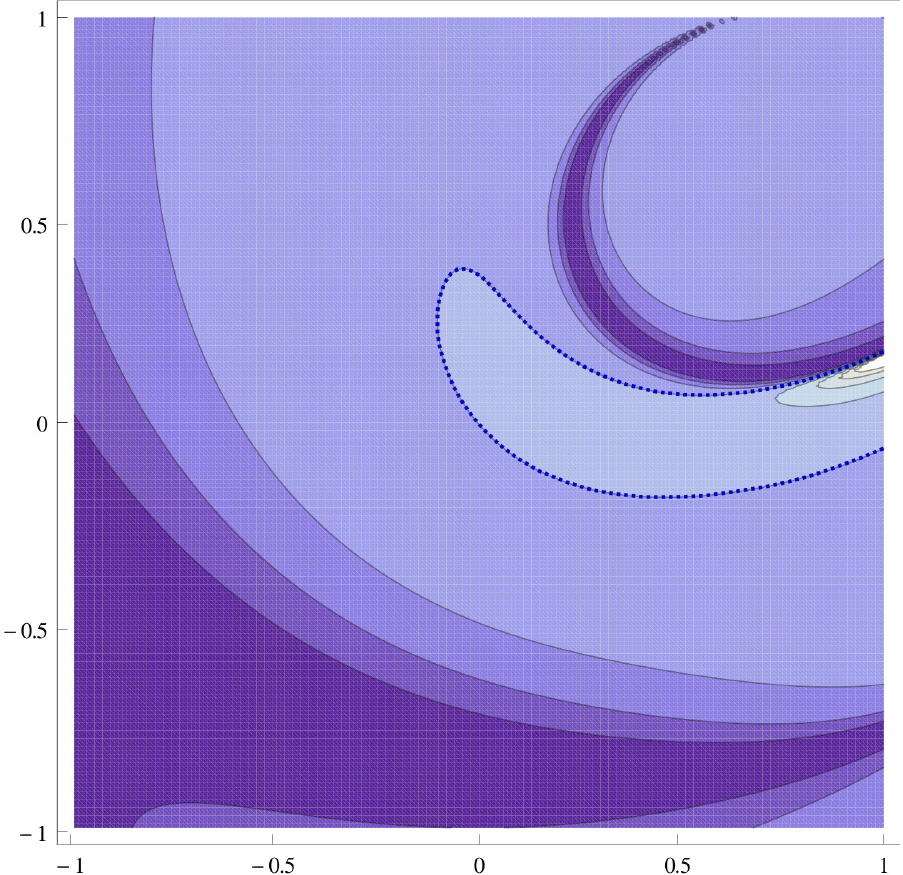} & \includegraphics[scale=1.1,keepaspectratio=true]{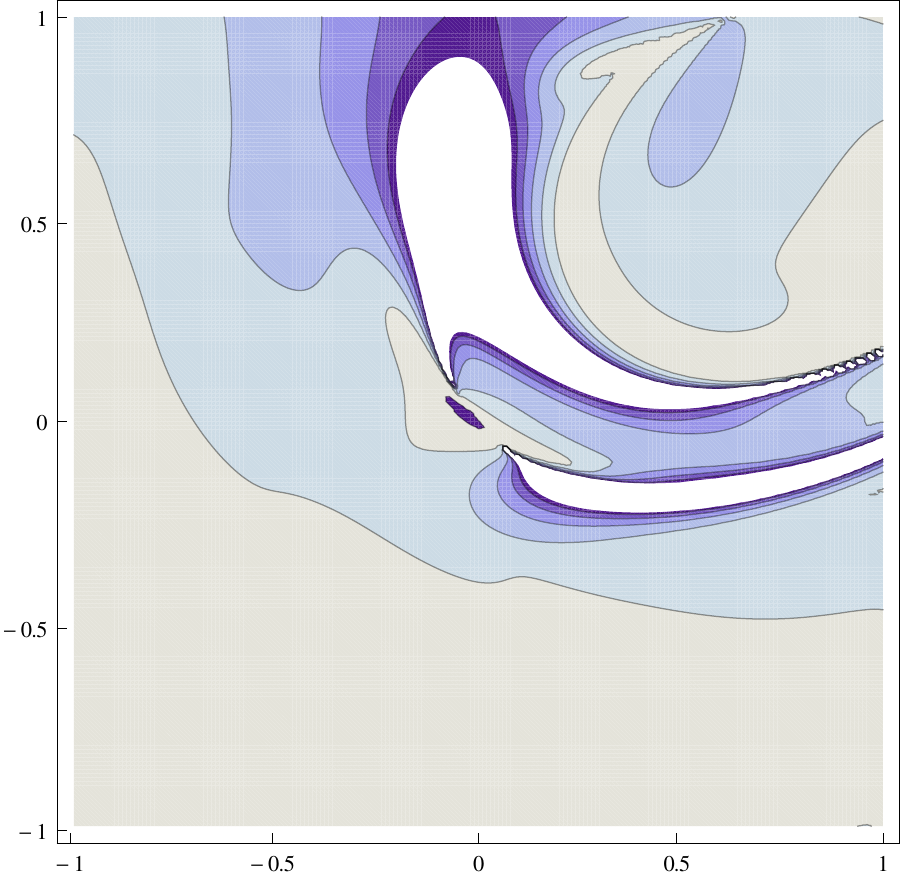} & \includegraphics[scale=1.1,keepaspectratio=true]{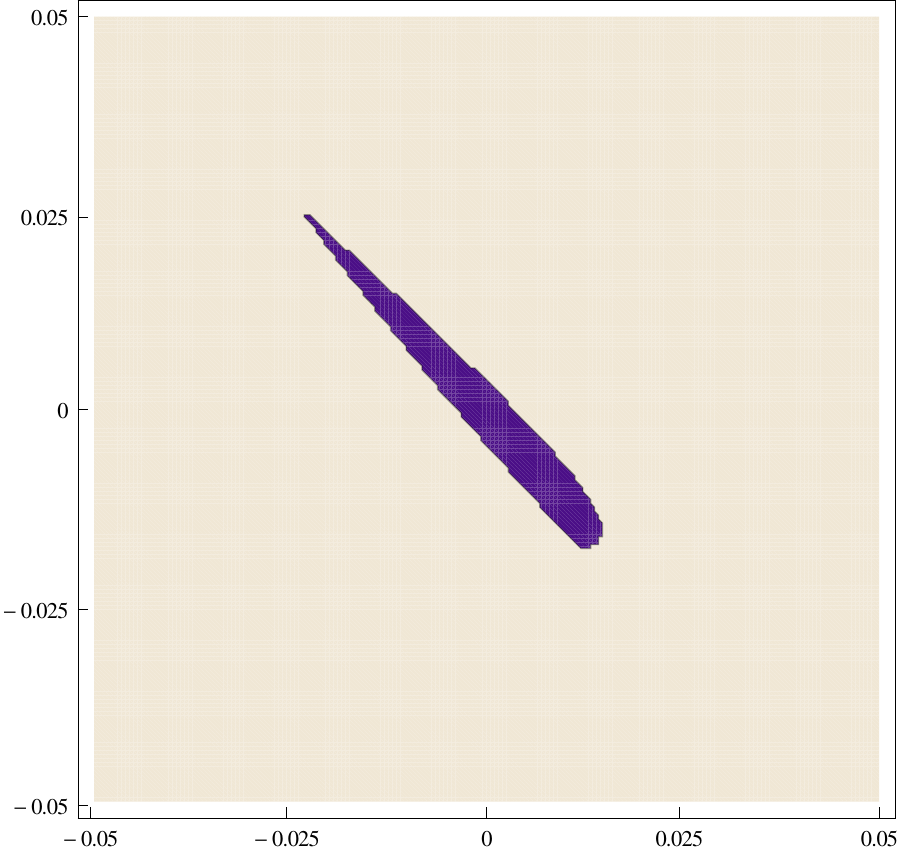}
\end{tabular}
}
\caption{{\it The parameter space of solutions around Sol.~1, 2 and 3 (top, center and bottom), projected on the $\left(A_{S},B_{S}\right)$ plane. The origin in each picture is the found stable dS. \emph{Left:} Level curves of the cosmological constant; the dS regions are the ones filled with lighter colours 
next to the Minkowski (blue dashed) lines.  
\emph{Middle:} Level curves of the $\eta$ parameter; the tachyon-free regions are filled with darker colours.
\emph{Right:} The tiny region of overlap in parameter space corresponding with stable dS is zoomed in here.}}
\label{fig:stable_dS}
\end{center}
\end{figure}

For each of the $3$ stable dS extrema, we have plotted the level curves of $\tilde{\g}\,\equiv\,\frac{|DW|^{2}}{3|W|^{2}}$ and of $\eta \,\equiv \,\textrm{Min}\left\{\textrm{Eigenvalues}\left({\left(m^{2}\right)^{I}}_{J}\right)\right\}$
in the $(A_{S},B_{S})$ plane. Subsequently we found the regions where $\tilde{\g}>1$ and $\eta>0$, corresponding to stable dS critical points. 
These results are depicted in figure~\ref{fig:stable_dS} for the three solutions.

Our present understanding of the plots suggests that stable dS vacua organise themselves into thin sheets in parameter space. This conclusion seems to be perfectly in line with what was found in 
ref.~\cite{Danielsson:2012by} in the isotropic case where only $N=6$ real fields were retained in the model.

\section*{Conclusions}

We have in this work presented three de Sitter vacua which are stable in all of the $N=14$ considered non-isotropic directions. These de Sitter vacua were obtained using the framework of the $STU$-models, more specifically $\mathbb{Z}_2 \times \mathbb{Z}_2$ orbifold compactifications with generalised fluxes. We used a setup containing $N=14$ real fields, with $N=14$ supersymmetry-breaking parameters and $2N=28$ fluxes; $F_3$, $H$ and non-geometric flux $Q$. This result shows that the addition of non-geometric flux provides enough freedom to allow for stable de Sitter solutions.

The details of the de Sitter solutions, the size of the cosmological constant and the values of the masses can be found in table~\ref{dS_sols1}. Eventough all masses are non-tachyonic there are two cases where the masses are very small compared to the value of the potential -- which might indicate the possibility of an instability. However this is not a general feature of these solutions, since only two out of three solutions contain such low masses.  We have also included table~\ref{dS_sols2} for the reader interested in the details of the values of the fluxes. These solutions were obtained via an evolutionary algorithm designed to seek out stable de Sitter regions. We have included a description of our implementation of this algorithm in appendix~\ref{app:algorithm}.

As expected by ref.~\cite{Danielsson:2012by}, these solutions are part of stable de Sitter regions that are organised into sheets or small regions - an overlap between regions with positive cosmological constant and stability - as can be seen in $2$-dimensional slices of the $N=14$ dimensional parameter space. We have included plots of these regions in the vicinity of the found stable de Sitter solutions in figure~\ref{fig:stable_dS}. The structure and occurrence of these overlapping regions is very fascinating and is something we plan to investigate further in the future.

Another very interesting question is whether the supergravity approximation is valid when the non-geometric fluxes are added. Because of the absence of a full description of the non-geometric fluxes in $10$D it is hard to believe that it would be. However as a future investigation we plan to study the supergravity approximation in terms of $4$D supergravity where non-geometric fluxes could be an acceptable addition.

%
%

\section*{Acknowledgments}

We would like to thank Kristoffer Ekman W\"arja for discussions regarding evolutionary algorithms. The work of the authors is supported by the Swedish Research Council (VR), and the G\"oran Gustafsson Foundation.

\newpage

%
%

\appendix

\section{Flux values and figures}
\label{app:fluxfig}

\begin{table}[h!]
\renewcommand{\arraystretch}{1.25}
\begin{center}
\scalebox{0.75}[0.83]{
\hspace{-10mm}
\begin{tabular}{ | c || c | c | c | }
\hline
  &  Sol.~$1$ & Sol.~$2$ & Sol.~$3$ \\
\hline 
\hline
$\begin{array}{cc} A_{S} & B_{S} \\ A_{T_{1}} & B_{T_{1}} \\ A_{T_{2}} & B_{T_{2}} \\ A_{T_{3}} & B_{T_{3}} \\ A_{U_{1}} & B_{U_{1}} \\ A_{U_{2}} & B_{U_{2}} \\ A_{U_{3}} & B_{U_{3}} \end{array}$
& $\begin{array}{cc} -0.314252 & -0.677296 \\ -0.145819 & 0.201488 \\ 0.324296 & 0.200242 \\ -0.510199 & -0.830019 \\ 0.437796 & -0.223916 \\ -0.10521 & -0.716205 \\ -0.110376 & -0.952221 \end{array}$ & 
$\begin{array}{cc} 0.266259 & 0.0513389 \\ 0.0113161 & 1.19326 \\ -0.374583 & -0.318198 \\ 0.0745176 & 0.814969 \\ -0.410472 & 0.56115 \\ -0.074538 & 0.276484 \\ -0.597945 & 0.903104 \end{array}$ & 
$\begin{array}{cc} -0.586674 & -0.186345 \\ 0.684213 & 0.846619 \\ 0.540869 & 1.10608 \\ -0.373167 & 0.724434 \\ -0.265422 & 0.187632 \\ -0.118099 & 0.651658 \\ -0.297359 & 0.75668 \end{array}$\\
\hline
\hline
$\begin{array}{cc} a_{0} & b_{0} \end{array}$ & $\begin{array}{cc} 0.249727 & 1.26278 \end{array}$  &  $\begin{array}{cc} -0.0809612 & -0.470688 \end{array}$ & $\begin{array}{cc} -0.126241 & 0.505546 \end{array}$ \\
\hline
$\begin{array}{cc} a_{1}^{(i)} & b_{1}^{(i)} \end{array}$ & $\begin{array}{cc} -0.385558 & -1.00688 \\ 3.24742 & 3.23091 \\ -3.42277 & -1.65416 \end{array}$  & 
$\begin{array}{cc} 0.763615 & 0.637699 \\ -0.129007 & -1.32755 \\ -0.258342 & -0.502742 \end{array}$ & 
$\begin{array}{cc} 0.0845892 & -0.342897 \\ -0.00376873 & 0.431055 \\ -0.687791 & -1.81103 \end{array}$ \\
\hline
$\begin{array}{cc} a_{2}^{(i)} & b_{2}^{(i)} \end{array}$ & $\begin{array}{cc} -0.678952 & 0.269654 \\ 4.77196 & -2.82032 \\ -4.74382 & 3.85504 \end{array}$  & 
$\begin{array}{cc} -1.69969 & -0.585701 \\ 0.141055 & -0.0290134 \\ 2.04863 & 0.623729 \end{array}$ & 
$\begin{array}{cc} -0.748501 & -0.346495 \\ 2.79635 & -0.40546 \\ -2.13249 & 0.457822 \end{array}$ \\
\hline
$\begin{array}{cc} a_{3} & b_{3} \end{array}$ & $\begin{array}{cc} -0.626634 & 0.244097  \end{array}$ & $\begin{array}{cc} -0.0281293 & -0.0877685 \end{array}$ & $\begin{array}{cc} -0.446200 & -0.270327 \end{array}$ \\ 
\hline
$c_{0}^{(i)}$ & $\begin{array}{c} 0.210031 \\ 0.680146 \\ -0.154349 \end{array}$ & $\begin{array}{c} 0.224759 \\ -0.16114 \\ 0.287961 \end{array}$ & 
$\begin{array}{c} 0.471208 \\ 0.327864 \\ -0.586172 \end{array}$\\
\hline
$c_{1}^{(ij)}$ & $\begin{array}{ccc} 0.934579 & 0.932678 & 0.134729 \\ 1.3931 & 0.978188 & 0.351523 \\ -1.12313 & -0.707561 & -0.313208 \end{array}$ & 
$\begin{array}{ccc} 1.00141 & -1.58335 & -2.92823 \\ -0.335711 & -0.460028 & 0.806899 \\ -0.628606 & 0.56901 & 1.78014 \end{array}$ & 
$\begin{array}{ccc} -0.108522 & 0.144519 & -1.70585 \\ 0.178853 & -0.083869 & 1.47008 \\ -0.489907 & -0.220766 & -0.305991 \end{array}$ \\
\hline
\end{tabular}
}
\end{center}
\caption{{\it The $3$ stable dS extrema found through the algorithm presented in appendix~\ref{app:algorithm}. The first part of the table shows the values of the $14$ SUSY-breaking parameters, 
whereas in the second part we give the explicit values of the $28$ fluxes turned on at the corresponding critical points.}}
\label{dS_sols2}
\end{table}

\newpage

\section{The algorithm}
\label{app:algorithm}
\numberwithin{equation}{section}

The search algorithm used is inspired by the methods of \emph{differential evolution}. These are evolutionary algorithms that can walk in a solution landscape to find extremal points or special regions. Here we will describe the how the algorithm works and parts of the workings of its implementation. We will start by presenting the steps of the algorithm, then proceed to explain them in closer detail.

The algorithm proceeds according to the following repeated scheme.
\begin{center}
\begin{tabular}{rl}
1. &  Generate a population (set of parents).\label{item:gen}\\
2. &  Create a mutated copy of the population (set of children).\label{item:mutate}\\
3. &  Generate $\tilde{\gamma}$ and the mass eigenvalues for parents and children.\label{item:solve}\\
4. &  For each parent and child pair, choose the \emph{better} one.\label{item:fight}\\
5. &  Check additional criteria and repeat until termination conditions.\label{item:terminate}
\end{tabular}
\end{center}

{\bf Step 1.} The system is determined when the $14$ SUSY-breaking parameters are specified. The population is a set where each element contain values for each of the $14$ SUSY-breaking parameters. This set can either be generated randomly, or by a particular choice. We chose to leave one parameter to be fixed by demanding that we start at a Minkowski point. The size of the population is a weigh-off against the amount of iterations one wants to get to in a certain time. A large population gives a higher chance of a better initial position in the parameter space, but more iterations would more likely converge into a de Sitter region.

{\bf Step 2.} Performing a mutation is to take a step in some direction. This direction is random and so is the step size. This gives the algorithm the possibility to search freely without us imposing any explicit directions or step sizes. However, the randomness of the mutation should be chosen carefully - a mutation too random will make the algorithm a random-walk instead of being evolutionarily progressed. This is why the following code was chosen to mutate a member of the population

\begin{code}
	mutate[member0_, sigma0_] := Module[{member=member0, sigma=sigma0},
	  Do[
	    If[Not[RandomInteger[{1, 3}] == 1],
	      member[[i]] = member[[i]] + RandomVariate[NormalDistribution[0, sigma]]
	    ];
	  ,{i, Length[member]}];
	  member
	];
\end{code}

The step direction is random in the sense that there is a two in three chance that a number will be mutated. The size of the step is also random according to a normal distribution with zero mean. The standard deviation is given as input to this module, so that the algorithm can for example increase the effective step size if the solution gets stuck. There are hence two parameters we need to give to the algorithm that will affect its efficiency, \emph{i.e.} the chance of getting mutated, and standard deviation for the mutation.

{\bf Step 3.} For each member we need the $\tilde{\gamma}$ and the mass matrix, to later be able to compare solutions. A module is needed to solve the set of equations and produce these parameters. This is where most of the computational time is spent; given the SUSY-breaking parameters one needs to solve $N$ linear equations and compute the eigenvalues of a $N\times N$ matrix. This means that it is preferable to associate the $\tilde{\gamma}$ and each mass with every member of the population, to avoid calculating these more than once.

{\bf Step 4.} Some function must be invented to decide whether the parent or child is the better solution. The choice made here was to have the following three criteria
\begin{center}
\begin{tabular}{rll}
1. & Approaching range & $\mathcal{N}(N^1,\sigma)$ ,\\
2. & Getting within range  & $\mathcal{N}(N^2,\sigma)$ ,\\
3. & Getting within numerically safe range & $\mathcal{N}(N^3,\sigma)$ ,
\end{tabular}
\end{center}
where $\mathcal{N}(\m,\sigma)$ denotes a random number chosen according to the normal distribution around the mean $\m$ with standard deviation $\sigma$.
This applies to each of the $15$ parameters that specify a stable dS extremum when all within range, \emph{i.e.} $\tilde{\g}$ plus the eigenvalues of the mass matrix.
All of the above conditions receive a number, its magnitude depending on how desirable the condition is. The total value parametrising the solution's desirability is given by the algebraic
sum of the $15$ separate contributions. This parameter will judge whether or not the parent should be replaced by its mutated child.
If one of the above criteria is exclusively not satisfied, it receives the corresponding negative number instead. In choosing this evaluation criterion, some limit cases should be considered - what to do in the case of one parameter approaching range and one getting further away, etc.. 
Our choice of the normal distribution with mean $N^p$, where the power $p$ categorises the conditions by desirability is to make random choices in such conflicting cases. This is hence another parameter we need to give as input to the algorithm.

{\bf Step 5.} There are also additional conditions that could be checked for the code being able to progress smoothly for an indefinite time or until max iterations. Such a condition could be how long a certain solution have been iterated without finding a better child. If the solution has stayed in the same position for too long, one might want to increase the size of the mutation (the standard deviation), to jump out of this position where it seemed not able to find stable de Sitter. This gives two parameters to input into the algorithm: how long before a solution is categorized as stuck and how much the standard deviation should increase. One could also imagine that if the solution is very close of having $\tilde{\gamma}>1$ and all non-negative masses, that one would decrease the size of the mutation to zoom in on a smaller search area, however this was not implemented by us.

An additional check is to keep track of how long after a solution got stuck this should be removed and replaced. Since the search area is increasing after getting stuck, the chance of finding a better solution within reasonable time decreases. These solutions can be saved but removed from the population, to be able to study them later if that is desired and to ensure the progression of the population. This gives another parameter as input to the algorithm. Similarly one should remove any stable de Sitter found, from the population and save it, this to make the population continue with the possibility to find more stable de Sitters.

A termination condition should be given for the loop. The number of iterations, together with the populations size, is a comfortable way to determine the time for the code to run.\\

As mentioned before, the algorithm has a number of parameters that needs to be specified. What follows are the values for these numbers which successfully produced stable de Sitter solutions.
\begin{center}
\begin{tabular}{ll}
Population size & $16$\\
Standard deviation for mutation & $\sigma_m = 0.01$\\
Chance to mutate a number & two in three\\
Condition desirability & $\mathcal{N}(N^p,\sigma)$ ($\sigma = 1$)\\
Iterations & $50\,000$\\
Iterations without change until stuck & 500\\
Growth of mutation standard deviation per iteration after stuck & $.5 \sigma_m$\\
Iterations without change until removed from population & 2000
\end{tabular}
\end{center}

%
%

\small

\bibliography{references}
\bibliographystyle{utphys}

\end{document}